\documentclass[11pt]{article}
\usepackage[a4paper,hmargin=1in,vmargin=1.3in]{geometry}
\usepackage[utf8]{inputenc}
\usepackage[english]{babel}
\usepackage{xcolor}
\usepackage{amssymb}
\usepackage{amsmath}
\usepackage{setspace}
\usepackage{tikz}
\usepackage{hyperref}
\usetikzlibrary{fit,graphs,graphs.standard,quotes,shapes,arrows}
\newcommand{\pro}{\noindent\textbf{Proof.}\ }
\newcommand{\qed}{\hspace*{\fill}\rule{1ex}{1ex}\vspace{1em}}
\newtheorem{theorem}{Theorem}
\newtheorem{lemma}{Lemma}

\onehalfspace

\begin{document}
\begin{center}

{\Large\bf Approximating the Maximum Number\\ of Synchronizing States in Automata} \vskip 2mm

{\large Andrew Ryzhikov} \vskip 2mm {\sl
United  Institute of Informatics Problems, National Academy of Sciences of Belarus, 6~Surganova Str., 220012, Minsk, Belarus, {\tt ryzhikov.andrew@gmail.com}
\vskip 2mm}
\end{center}

\begin{abstract} We consider the problem {\sc Max Sync Set} of finding a maximum synchronizing set of states in a given automaton. We show that the decision version of this problem is PSPACE-complete and investigate the approximability of {\sc Max Sync Set} for binary and weakly acyclic automata (an automaton is called weakly acyclic if it contains no cycles other than self-loops). We prove that, assuming $P \ne NP$, for any $\varepsilon > 0$, the {\sc Max Sync Set} problem cannot be approximated in polynomial time within a factor of $O(n^{1 - \varepsilon})$ for weakly acyclic $n$-state automata with alphabet of linear size, within a factor of $O(n^{\frac{1}{2} - \varepsilon})$ for binary $n$-state automata, and within a factor of $O(n^{\frac{1}{3} - \varepsilon})$ for binary weakly acyclic $n$-state automata. Finally, we prove that for unary automata the problem becomes solvable in polynomial time.
\end{abstract}


\section{Introduction}\label{sect-intro}

Let $A = (Q, \Sigma, \delta)$ be a deterministic finite automaton, where $Q$ is a set of states, $\Sigma$ is a finite alphabet, and $\delta: Q \times \Sigma \to Q$ is a transition function. Note that our definition does not include input and output states. Let $\Sigma^*$ be the set of all finite words over the alphabet $\Sigma$. The mapping $\delta$ can be extended in a natural way into a $Q \times \Sigma^* \to Q$ mapping as follows: we take $\delta(s, x w) = \delta(\delta(s, x), w)$ for each letter $x \in \Sigma$, state $s \in Q$, and non-empty word $w \in \Sigma^*$. An automaton $A = (Q, \Sigma, \delta)$ is called {\em synchronizing} if there exists a word $w$ that maps every state to a particular common state $q \in Q$, i.\,e. $\delta(s, w) = q$ for each $s \in Q$. An automaton is called {\em binary} if its alphabet has size two, and {\em unary} if it has size one. A {\em cycle} in an automaton is a sequence $q_1, \ldots, q_n$ of its states such that there exist letters $x_1, \ldots, x_n \in \Sigma$ with $\delta(q_i, x_i) = q_{i + 1}$ for $1 \le i \le n - 1$ and $\delta(q_n, x_n) = q_1$. A cycle is a {\em self-loop} if it consists of one state. An automaton is called {\em weakly acyclic} if all its cycles are self-loops. Weakly acyclic automata were explicitly introduced in \cite{Jiraskova2012} under the name of acyclic automata. We prefer the term weakly acyclic, as the term acyclic is usually used for automata recognizing finite languages \cite{Watson2003}. Earlier weakly acyclic automata were mentioned in connection with the problem of recognizing piecewise testable languages \cite{Simon1975, Stern1985}.

The concept of synchronization is widely studied in automata theory and has applications in robotics, biocomputing, semigroup theory and symbolic dynamics (see survey \cite{Volkov2008} and references therein). It is also a key notion in the famous \v{C}ern{\'y} conjecture about the length of the shortest synchronizing word in automata \cite{Cerny1971}. The problem of deciding whether a given automaton $A$ is synchronizing can be reduced to a reachability problem in an automaton build on pairs of states of the automaton $A$, and thus is solvable in polynomial time \cite{Volkov2008}. However, the problem of finding the shortest synchronizing word for binary automata is hard to approximate \cite{Gawrychowski2015}.

A set $S \subseteq Q$ of states in an automaton $A$ is called {\em synchronizing} if there exists a word $w \in \Sigma^*$ and a state $q \in Q$ such that the word $w$ maps each state $s \in S$ to the state $q$. The word $w$ is said to {\em synchronize} the set $S$. It follows from the definition that an automaton is synchronizing if and only if the set $Q$ of all its states is synchronizing. Consider the problem {\sc Sync Set} of deciding whether a given set $S$ of states of a given automaton $A$ is synchronizing.

\vspace{.25cm}

{\sc Sync Set}

{\em Input}: An automaton $A$ and a subset $S$ of its states;

{\em Output}: Decide whether $S$ is a synchronizing set in $A$.

\vspace{.25cm}

The {\sc Sync Set} problem is PSPACE-complete \cite{Rystsov1983}, \cite{Sandberg2005}, even for strongly connected binary automata \cite{Vorel2014}, \cite{VorelThesis2015}. Its motivation is the following: assume that the current state of an automaton $A$ is unknown and cannot be observed, but it is known to belong to a given subset of states $S$. We know the transition function of $A$. Can we map all the states of the automaton $A$ by some word to a particular state, thus resolving the initial state uncertainty? In this paper, we consider a related {\sc Max Sync Set} problem, which is to find a maximum cardinality set of states in an automaton such that the initial state uncertainty can be resolved for it.

\vspace{.25cm}

{\sc Max Sync Set}

{\em Input}: An automaton $A$;

{\em Output}: A synchronizing set of states of maximum size in $A$.

\vspace{.25cm}

T\"{u}rker and Yenig\"{u}n \cite{Turker2015} study a variation of this problem, which is to find a set of states of maximum size that can be mapped by some word to a subset of a given set of states in a given monotonic automaton. They reduce the {\sc N-Queens Puzzle} problem \cite{Bell2009} to this problem to prove its NP-hardness. However, their proof is not correct, as their reduction is not polynomial: the input has size $O(\log N)$, and the output size is polynomial in $N$.

We assume that the reader is familiar with the notions of an approximation algorithm and a gap-preserving reduction (for reference, see the book by Vazirani~\cite{Vazirani2001}), and PSPACE-completeness (refer to the book by Sipser~\cite{Sipser2006}). We shall also need some results from Graph Theory. An {\em independent set} $I$ in a graph $G$ is a set of its vertices such that no two vertices in $I$ share an edge. The size of a maximum independent set in $G$ is denoted $\alpha(G)$. The {\sc Independent Set} problem is defined as follows.

\vspace{.25cm}

{\sc Independent Set}

{\em Input}: A graph $G$;

{\em Output}: An independent set of maximum size in $G$.

\vspace{.25cm}

Zuckerman \cite{Zuckerman2006} has proved that, unless P = NP, there is no polynomial \mbox{$p^{1 - \varepsilon}$-approximation} algorithm for the {\sc Independent Set} problem for any $\varepsilon > 0$, where $p$ is the number of vertices in a given graph.

In this paper, we show that the decision version of the {\sc Max Sync Set} problem is PSPACE-complete for binary automata. We prove that, unless P = NP, for any $\varepsilon > 0$, the size of a maximum synchronizing set in a given $n$-state automaton cannot be approximated in polynomial time within a factor of $O(n^{1-\varepsilon})$ for weakly acyclic automata, within a factor of $O(n^{\frac{1}{2} - \varepsilon})$ for binary automata, and within a factor of $O(n^{\frac{1}{3} - \varepsilon})$ for binary weakly acyclic automata. We also show that for unary automata, the {\sc Max Sync Set} problem is solvable in polynomial time.

\section{The {\sc Max Sync Set} Problem}

First we investigate the PSPACE-completeness of the decision version of the  {\sc Max Sync Set} problem, which we shall denote as {\sc Max Sync Set-D}. Its formulation is the following: given an automaton $A$ and a number $c$, decide whether there is a synchronizing set of states of cardinality at least $c$ in $A$.

\begin{theorem}\label{thm-pspace-general}
	The {\sc Max Sync Set-D} problem is PSPACE-complete for binary automata.
\end{theorem}
\pro The {\sc Sync Set} problem is in PSPACE \cite{Sandberg2005}. Thus, the {\sc Max Sync Set-D} problem is also in PSPACE, as we can sequentially check whether each subset of states is synchronizing and compare the size of a maximum synchronizing state to $c$.

To prove that the {\sc Max Sync Set-D} problem is PSPACE-hard for binary automata, we shall reduce a PSPACE-complete {\sc Sync Set} problem for binary automata to it \cite{Vorel2014}. Let an automaton $A$ and a subset $S$ of its states be an input to {\sc Sync Set}. Let $n$ be the number of states of $A$. Construct a new automaton $A'$ by initially taking a copy of $A$. For each state $s \in S$, add $n + 1$ {\em new} states to $A'$ and define all the transitions from these new states to map to $s$, regardless of the input letter. Define the set $S'$ to be a union of all new states and take $c = |S'| = (n + 1)|S|$.

Let $S_1$ be a maximum synchronizing set in $A$ not containing at least one new state $q$. As $S_1$ is maximum, it does not contain other $n$ new states that can be mapped to the same state as $q$. Thus, the size of $S_1$ is at most $n + (n + 1)|S| - (n + 1) < (n + 1)|S| = c$. Hence, each synchronizing set of size at least $c$ in $A'$ contains $S'$. The set $S$ is synchronizing in $A$ if and only if $S'$ is synchronizing in $A'$, as each word $w$ synchronizing $S$ in $A$ corresponds to a word $xw$ synchronizing $S'$ in $A'$, where $x$ is an arbitrary letter. Thus, $A'$ has a synchronizing set of size at least $c$ if and only if $S$ is synchronizing in $A$. \qed

Now we proceed to inapproximability results for the {\sc Max Sync Set} problem in several classes of automata.

\begin{theorem}\label{thm-inapprox-alph}
	The problem {\sc Max Sync Set} for weakly acyclic $n$-state automata over an alphabet of cardinality $O(n)$ cannot be approximated in polynomial time within a factor of $O(n^{1 - \varepsilon})$ for any~$\varepsilon > 0$ unless P = NP.
\end{theorem}
\pro We shall prove this theorem by constructing a gap-preserving reduction from the {\sc Independent Set} problem. Given a graph $G = (V, E)$, $V = \{v_1, v_2, \ldots, v_p\}$, we construct an automaton $A = (Q, \Sigma, \delta)$ as follows. For each $v_i \in V$, we construct two states $s_i, t_i$ in $Q$. We also add a state $f$ in $Q$. Thus, $|Q| = 2p + 1$. The alphabet $\Sigma$ consists of letters $\tilde{v}_1, \ldots, \tilde{v}_p$ corresponding to the vertices of $G$.

The transition function $\delta$ is defined in the following way. For each $1 \le i \le p$, the state $s_i$ is mapped to $f$ by the letter $\tilde{v}_i$. For each $v_iv_j \in E$ the state $s_i$ is mapped to $t_i$ by the letter $\tilde{v}_j$, and the state $s_j$ is mapped to $t_j$ by the letter $\tilde{v}_i$. All yet undefined transitions map a state to itself.

\begin{lemma} Let $I$ be a maximum independent set in $G$. Then the set $S = \{s_i \mid v_i \in I\} \cup \{f\}$ is a synchronizing set of maximum cardinality (of size $\alpha(G) + 1$) in the automaton $A = (Q, \Sigma, \delta)$. 
\end{lemma}
\pro Let $w$ be a word obtained by concatenating the letters corresponding to $I$ in arbitrary order. Then $w$ synchronizes the set $S = \{s_i \mid v_i \in I \} \cup \{f\}$ of states of cardinality $|I| + 1$. Thus, $A$ has a synchronizing set of size at least $\alpha(G) + 1$.

In other direction, let $w$ be a word synchronizing a set of states $S'$ of maximum size in $A$. We can assume that after reading $w$ all the states in $S'$ are mapped to $f$, as all the sets of states that are mapped to any other state have cardinality at most two. Then by construction there are no edges in $G$ between any pair of vertices in $I' = \{v_i \mid s_i \in S'\}$, so $I'$ is an independent set of size $|S'| - 1$ in $G$. Thus the maximum size of a synchronizing set in $A$ equals to $\alpha(G) + 1$. \qed

Thus we have a gap-preserving reduction from the {\sc Independent Set} problem to the {\sc Max Sync Set} problem with a gap $\Theta(p^{1 - \varepsilon})$ for any $\varepsilon > 0$. It is easy to see that $n = \Theta(p)$ and $A$ is weakly acyclic, which concludes the proof of the theorem. \qed

Next we move to a slightly weaker inapproximability result for binary automata.

\begin{theorem}\label{thm-inapprox-gen}
	The problem {\sc Max Sync Set} for binary $n$-state automata cannot be approximated in polynomial time within a factor of $O(n^{\frac{1}{2} - \varepsilon})$ for any $\varepsilon > 0$ unless P = NP.
\end{theorem}
\pro Again, we construct a gap-preserving reduction from the {\sc Independent Set} problem similar to the proof of Theorem \ref{thm-inapprox-alph}. Given a graph $G = (V, E), V = \{v_1, v_2, \ldots, v_p\}$, we construct an automaton $A =(Q, \Sigma, \delta)$ in the following way. Let $\Sigma = \{0, 1\}$. First we construct the main gadget $A_{main}$ having a synchronizing set of states of size $\alpha(G)$. For each vertex $v_i \in V, 1 \le i \le p$, we construct a set of new states $L_i = V_i \cup U_i$, where $V_i = \{v^{(i)}_j : 1 \le j \le p\}, U_i = \{u^{(i)}_j : 1 \le j \le p\}$, in~$Q$. We call $L_i$ the $i$th {\em layer} of $A_{main}$. We also add a state $f$ to $Q$. For each $i$, $1 \le i \le p$, the transition function $\delta$ is defined as:
 
\[ \delta(v^{(i)}_j, 0) = \left\{ 
\begin{array}{l l}
u^{(i)}_j & \quad \mbox{if $i = j$,}\\
v^{(i + 1)}_j & \quad \mbox{otherwise}\\
\end{array} \right. \]

\[ \delta(v^{(i)}_j, 1) = \left\{ 
\begin{array}{l l}
u^{(i)}_j & \quad \mbox{if there is an edge $v_i v_j \in E$,}\\
v^{(i + 1)}_j & \quad \mbox{otherwise}\\
\end{array} \right. \]

Here all $v^{(n + 1)}_j, 1 \le j \le p$, coincide with $f$. For each state $u^{(i)}_j$, the transitions for both letters $0$ and $1$ lead to the originating state (i.e. they are self-loops).

We also add an $p$-state cycle $A_{cycle}$ attached to $f$. It is a set of $p$ states $c_1, \ldots, c_p$, mapping $c_i$ to $c_{i + 1}$ and $c_p$ to $c_1$ regardless of the input symbol. Finally, we set $c_1$  to coincide with $f$. Thus we get the automaton $A_1$. Figure \ref{fig:exampleA} presents an example of $A_1$ for a graph with three vertices $v_1, v_2, v_3$ and one edge~$v_2v_3$.

\setlength{\unitlength}{2.2pt}
\begin{figure}[hbt]
\begin{center}

\begin{tikzpicture}[->,>=latex',
vertex/.style={circle, draw=black, fill, minimum width=1.5mm, inner sep=0pt, outer sep=0pt},
every label/.style={inner sep=0pt, minimum width=0pt, label distance=0.1mm},
yscale=-2.5,
xscale=2.5
]
\graph[nodes=vertex, empty nodes, no placement] {
	{
		v11[x=0,y=0,label=above:$v^{(1)}_1$] -> [edge label=1]
		v12[x=1,y=0,label=above:$v^{(2)}_1$] -> [edge  label={$0$,$1$}]
		v13[x=2,y=0,label=above:$v^{(3)}_1$] -> [edge  label={$0$,$1$}]
		f[x=3,y=1,label=above:$f$]
	};
	{
		v11 -> [edge label=0, swap]	u11[x=0.25,y=0.5,label=right:$u^{(1)}_1$]
	};
	{
		v21[x=0,y=1,label=above:$v^{(1)}_2$] -> [edge  label={$0$,$1$}]
		v22[x=1,y=1,label=above:$v^{(2)}_2$] -> [edge label=1]
		v23[x=2,y=1,label=above:$v^{(3)}_2$] -> [edge label=0]
		f
	};
	{
		v22 -> [edge label=0]	u22[x=1.25,y=1.5,label=left:$u^{(2)}_2$]
	};
	{
		v31[x=0,y=2,label=above:$v^{(1)}_3$] -> [edge  label={$0$,$1$}]
		v32[x=1,y=2,label=above:$v^{(2)}_3$] -> [edge label=0]
		v33[x=2,y=2,label=above:$v^{(3)}_3$] -> [edge label=1,swap]
		f
	};
	{
		v32 -> [edge label=1,swap]	u32[x=1.25,y=2.5,label=right:$u^{(2)}_3$]
	};
	{
		v23 -> [edge label=1]	u23[x=2.25,y=1.5,label=left:$u^{(3)}_2$]
	};
	{
		v33 -> [edge label=0, swap]	u33[x=2.25,y=2.5,label=right:$u^{(3)}_3$]
	};
	{
		f -> [edge label={$0$,$1$},swap]
		c1[x=4,y=1.5,label=right:$c_2$] -> [edge label={$0$,$1$},swap]
		c2[x=4,y=0.5,label=right:$c_3$] -> [edge label={$0$,$1$},swap]
		f
	};
};
\node[rectangle,dotted,draw,fit=(v11)(u11)(v21)(v31),rounded corners=5mm,inner sep=22pt,label=below:$L_1$] {};

\end{tikzpicture}\textsl{}
\caption{An example of $A_1$. Unachievable states and self-loops are omitted.} \label{fig:exampleA}
\end{center}
\end{figure}
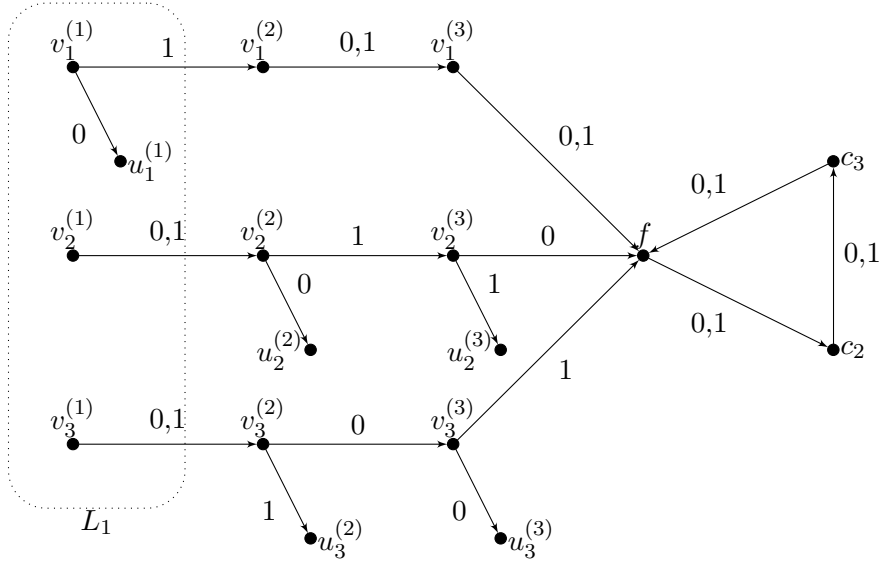
The main property of $A_1$ is claimed by the following lemma.

\begin{lemma}\label{lemma-main}
	The size of a maximum synchronizing set of states from the first layer in $A_1$ equals~$\alpha(G)$.
\end{lemma}
\pro Let $I$ be a maximum independent set in $G$. Consider a word $w$ of length $p$ such that its $i$th letter equals to $0$ if $v_i \notin I$ and to $1$ if $v_i \in I$. By the construction of $A_1$, this word synchronizes the set $\{v^{(1)}_j \mid v_j \in I\}$. Conversely, a synchronizing set of states from the first layer can be mapped only to some vertex of $A_{cycle}$, and the corresponding set of vertices in $G$ is an independent set. \qed

Some layer in the described construction can contain a synchronizing subset of size larger than the maximum synchronizing subset of the first layer. To avoid that, we modify $A_1$ by repeating each state (with all transitions) of the first layer $p$ times. More formally, we replace each pair of states $v^{(1)}_j$, $u^{(1)}_j$ with $p$ different pairs of states such that in each pair all the transitions repeat the transitions between $v^{(1)}_j$, $u^{(1)}_j$, and all the other states of the automaton. We denote the automaton thus constructed as $A$.

The following lemma claims that the described procedure of constructing $A$ from $G$ is a gap-preserving reduction from the {\sc Independent Set} problem in graphs to the {\sc Max Sync Set} problem in binary automata.

\begin{lemma}\label{lemma-copy}
	If $\alpha(G) > 1$, then the maximum size of a synchronizing set in $A$ equals~$n\alpha(G) + 1$.
\end{lemma}
\pro Note that due to the construction of $A_{cycle}$, each synchronizing set of $A$ is either a subset of a single layer of $A$ together with a state in $A_{cycle}$ or a subset of a set $\{v^{(i)}_j \mid 2 \le i \le \ell\} \cup \{u^{(\ell)}_j\}$ for some $\ell$ and $j$, together with $p$ new states that replaced $v^{(1)}_j$. Consider the first case. If some maximum synchronizing set $S$ contains a state from the $i$th layer of $A$ and $i > 1$, then its size is at most $p + 1$. The maximum synchronizing set containing some states from the first layer of $A$ consists of $p\alpha(G)$ states from this layer (according to Lemma \ref{lemma-main}) and some state of $A_{cycle}$, so this set has size $p\alpha(G) + 1 \ge 2p + 1$. In the second case, the maximum size of a synchronizing set is at most $p + (p - 1) + 1 = 2p < p \alpha(G) + 1$.
\qed

It is easy to see that the constructed reduction is gap-preserving with a gap $\Theta(p^{1 - \varepsilon}) = \Theta(n^{\frac{1}{2} - \varepsilon})$, where $n$ is the number of states in $A$, as $n = \Theta(p^2)$. Thus the {\sc Max Sync Set} for $n$-state binary automata cannot be approximated in polynomial time within a factor of $O(n^{\frac{1}{2} - \varepsilon})$ for any $\varepsilon > 0$ unless~P~=~NP, which concludes the proof of the theorem. \qed

Theorem \ref{thm-inapprox-gen} can also be proved by using Theorem \ref{thm-inapprox-alph} and a slight modification of the technique used in \cite{Vorel2014} for decreasing the size of the alphabet. However, in this case the resulting automaton is far from being weakly acyclic, while the automaton in the proof of Theorem \ref{thm-inapprox-alph} has only one cycle. The next theorem shows how to modify our technique to prove an inapproximability bound for {\sc Max Sync Set} in binary weakly acyclic automata.

\begin{theorem}\label{thm-inapprox-monot}
	The {\sc Max Sync Set} problem for binary weakly acyclic $n$-state automata cannot be approximated in polynomial time within a factor of $O(n^{\frac{1}{3} - \varepsilon})$ for any $\varepsilon > 0$ unless P = NP.
\end{theorem}
\pro We modify the construction of the automaton $A_{main}$ from Theorem \ref{thm-inapprox-gen} in the following way. We repeat each state (with all transitions) of the first layer $p^2$ times in the same way as it is done in the proof of Theorem \ref{thm-inapprox-gen}. Thus we get a weakly acyclic automaton $A_{wa}$ with $n = \Theta(p^3)$ states, where $p$ is the number of vertices in the graph $G$. Furthermore, similar to Lemma \ref{lemma-copy}, the size of the maximum synchronizing set of states in $A_{wa}$ is between $p^2\alpha(G)$ and $p^2\alpha(G) + p(p-1) + 1$, because some of the states from the layers other than the first can be also mapped to $f$. Both of the values are of order $\Theta(p^2\alpha(G))$, thus we have an gap-preserving reduction providing the inapproximability within a factor of $O(p^{1 - \varepsilon}) = O(n^{\frac{1}{3} - \varepsilon})$ for any $\varepsilon > 0$, where $n$ is the number of states in $A_{wa}$.\qed

We finish by noting that for unary automata the {\sc Max Sync Set} problem is solvable in polynomial time.

\begin{theorem}\label{thm-polytime-unary}
	The problem {\sc Max Sync Set} can be solved in polynomial time for unary automata.
\end{theorem}
\pro Consider the digraph $G$ induced by states and transitions of an unary automaton $A$. By definition, each vertex of $G$ has outdegree $1$. Thus, the set of the vertices of $G$ can be partitioned into directed cycles and a set of vertices not belonging to any cycle, but lying on a directed path leading to some cycle. Let $n$ be the number of states in $A$. It is easy to see that after performing $n$ transitions, each state of $A$ is mapped into a state in some cycle, and all further transitions will not map any two different states to the same state. Thus, it is enough to perform $n$ transitions and select such state $s$ that the maximum number of states are mapped to $s$.\qed

\section{Conclusions and open problems}

In this paper we have considered the problem of finding a maximum size synchronizing set in a given automaton. We showed that its decision version is PSPACE-complete. We proved that, unless P = NP, this problem cannot be approximated in polynomial time within a factor of, respectively, $O(n^{1 - \varepsilon})$, $O(n^{\frac{1}{2} - \varepsilon})$ and $O(n^{\frac{1}{3} - \varepsilon})$ for any $\varepsilon > 0$ for weakly acyclic, binary and binary weakly acyclic automata with $n$ states. For unary automata, we have shown that the {\sc Max Sync Set} problem is solvable in polynomial time.

A natural open question is the complexity of the {\sc Sync Set} problem for weakly acyclic and binary weakly acyclic automata. Another direction of study is the improvement of the presented inapproximability bounds and the development of approximation algorithms for the considered problems. It is unclear for us whether even an $O\left(\frac{n}{\log n}\right)$-approximation algorithm exists for $n$-state automata, by analogy with the {\sc Independent Set} problem approximability \cite{Boppana1992}. It is also interesting to investigate the complexity of {\sc Max Sync Set} for other classes of automata, such as monotonic, one-cluster, strongly connected, circular, and so on \cite{VorelThesis2015}.

\section{Acknowledgements}

We thank Peter Cameron for introducing us to the notion of synchronizing automata, and Vojt\v{e}ch Vorel and Yury Kartynnik for very useful discussions.

{\footnotesize
\bibliography{SyncBib}
\bibliographystyle{alpha}}

\end{document}